\documentclass[doublespacing]{elsart}

\usepackage{graphicx}

\usepackage{amssymb}
\journal{Statistical Mechanics Research. Nova}
\begin{document}

\begin{frontmatter}

\title{Anomalous diffusion coefficient in disordered media
from NMR relaxation}

\author{A.E. Sitnitsky},
\ead{sitnitsky@mail.knc.ru}

\address{Institute of Biochemistry and Biophysics, P.O.B. 30, Kazan
420111, Russia. e-mail: sitnitsky@mail.knc.ru }

\begin{abstract}
Application of fractional calculus to the description of anomalous
 diffusion and relaxation processes in complex media provided one of
 the most impressive impulses to the development of statistical
 physics during the last decade. In particular the so-called
fractional diffusion equation enabled one to capture the main
features of anomalous diffusion. However the price for this
 achievement is rather high - the fractional diffusion coefficient
 becomes an involved function of a characteristic of the media
(e.g., that of the radius of pores in the case of the porous one).
 Revealing this dependence from the first principles is one of the
 main problems in this field of science. Another one still remains
that of extracting this dependence from the experiment. The latter
 problem is tackled in the present paper. Our aim is to provide
 detailed and pedagogical deriving the relationship of the fractional
 diffusion coefficient with experimentally observable value from
nuclear magnetic resonance (NMR) spin-lattice relaxation data.
 The result obtained promotes the NMR
 relaxation method to become a powerful tool in solving the problem
of experimental measuring the fractional diffusion coefficient. Also
the merits and limitations of NMR relaxation method and pulsed-field gradient (PFG)
NMR for the research of anomalous diffusion are compared and discussed.
\end{abstract}

\end{frontmatter}

\section{Introduction}
Revealing the origin and adequate theoretical description
 of diffusion and relaxation processes have been one of the
 main concerns for statistical physics from the very beginning
 of this field of science. The obstacles that prevent achieving
these aims are augmented substantially with the increase of the
complexity of the system under consideration. That is why the
trend to research these processes in the so called complex
systems (for review see \cite{Wes03}, \cite{Met00}, \cite{Met04}
and refs. therein)
 posed new challenges for statistical physics and gave new
 impetus to its development. The history of this trend
goes back as far as to the middle of the 19-th century when
Kolraush introduced his
famous streched exponential function for the description of charge
relaxation in a Leyden gas. However only the end of
 the 20-th century is marked by a real breakthrough and by an
explosion of activity in this field of science.

Fractional dynamics is a modern and fruitful approach to the
description of anomalous diffusion processes in randomly
disordered media (see \cite{Met00}, \cite{Met04} and refs.
therein). The idea of this approach
goes back to many pioneers whose achievements are honored in the
reviews mentioned above. The reason for its
introducing is as follows. As is well known the mean squared
displacement of a free particle in a homogeneous media grows
linearly with time $<({\bf r}\ \ -<{\bf r}>)^2>=2dC_1t$ where
$C_1$ is the
diffusion coefficient with the dimension $cm^2/s$ and $d$ is the
(embedding) spatial dimension. This
conventional Einstein relationship of the classical theory results
from the ordinary diffusion equation for the probability density
function to find the particle at position x at time t and is a
direct consequence of the Fick's second law. However this simple
law does not take place in the inhomogeneous disordered media.
The generalization of the Fick's law for complex systems requires
conceptually new physical ideas and even envokes to new and rather unusual
for common audience mathematics - the so called fractional
calculus \cite{Ol74}, \cite{Mil93}, \cite{Pod98}, \cite{Wes03}.

 One of the basic tools of the fractional dynamics is a fractional
 diffusion equation (FDE) in which a fractional
 diffusion coefficient (FDC) $C_\alpha$ with an unusual dimension
$cm^2/s^{\alpha}$ is presented. The case of subdiffusion ($0 <
\alpha \leq 1$) is ubiquitous in nature yielding the mean squared
displacement $<({\bf r}\ \ -<{\bf r}>)^2>=2dC_{\alpha}t^{\alpha}$
 and it originates in
any fractal media due to the presence of dead ends on current
ways. The FDC acquires a functional dependence on the
phenomenological parameter $\alpha$ referring to the extent of
inhomogeneity of the system with $\alpha =1$ corresponding to the
case of ordinary diffusion. To establish this dependence for each
class of systems is one of the main problems for the theory.

On the other hand the problem has been posed to determine the FDC
experimentally. It was pioneered by \cite{Kl02} where the
interdiffusion of heavy and light water in a porous media was
observed by means of NMR. The authors made use of a one-
dimensional form of the FDE and found $C_\alpha$ using the
special case of its $\alpha =2/3$ solution. In \cite{Kos03},
\cite{Kos05} the procedure is called "neither very accurate nor
of general use". Though such critique may be justified for the
particular procedure of the paper \cite{Kl02} one should not
think that the abilities of NMR for measuring FDC are limited
somehow. From their own side the authors of \cite{Kos03},
\cite{Kos05} proposed a method to measure the FDC with the help
of a membrane system where the substance of interest is
transported in a solvent from one vessel to another
across a thin membrane. In our opinion namely their method is not
of general use because it requires the incorporation of a
membrane into the system of interest and thus can hardly be
applied to, e.g., porous materials. In
contrast NMR belongs to non-destructive and non-invasive methods
and thus can indeed be of general use. NMR diffusometry
\cite{Kim97} is a powerful method for investigation of
subdiffusion processes and it is widely applied to exploring
transport in porous \cite{Kim97}, \cite{Kim02}, \cite{Dul92},
percolative \cite{Kl97}, \cite{Kl99}, \cite{Kl02} and polymeric
\cite{Fi96} systems. The theory of NMR diffusometry in disordered
 media
\cite{Kim02}, \cite{Kl02}, \cite{Sit05} is developed within the
framework of fractional dynamics
\cite{Met00}, \cite{Met04}. The aim of this chapter is to show how
one can retrieve the functional dependence of the FDC on the
parameter $\alpha$ from the NMR spin-lattice relaxation data. We
obtain a simple analytical formula relating the FDC with the
contribution to the spin-lattice relaxation time by anomalous
translational diffusion that can in principle be extracted from
the experiment.

\section{Setting the stage}

As is well known in the theory of spin-lattice relaxation by
dipole-dipole interaction (see e.g. \cite{Ab61}) the contribution
to the spin-lattice relaxation rate constant due to translational
diffusion with the spectral density at a Larmor frequency $\omega
_{L}$ of the correlation function for spherical harmonics has the
form
\begin{equation}
\label{eq1}(1/T_{1})_{trans}=\frac{3\gamma^4\hbar^2I(I+1)}{2}
\{J^{(1)}(\omega _{L})+ J^{(2)}(2\omega _{L})\}
\end{equation}
where $\gamma$ is the gyromagnetic ratio of the nucleus, $I$ is
their spin and $\hbar$ is the Planck constant. The spectral
densities are proportional to the function $J(\omega)$ of the
spectral density
\begin{equation}
\label{eq2} J^{(1)}(\omega )=\frac{8\pi}{15} J(\omega )
\end{equation}
and
\begin{equation}
\label{eq3}J^{(2)}(\omega )=\frac{32\pi}{15} J(\omega )
\end{equation}
Thus the function $J(\omega)$ becomes the key object of the
analysis. The consideration of the spin-lattice relaxation in
homogeneous media due to diffusion motion of the particles
presented in \cite{Ab61} is based on the ordinary diffusion
equation for the probability density. To describe the case of
inhomogeneous and disordered (e.g., porous) media the
generalization of the ordinary diffusion equation within the
fractional calculus was suggested by Schneider and Wyss
\cite{Sc89} which includes the FDC $C_{\alpha }$ with the
dimension [cm$^{2}$/s$^{\alpha }$]. We introduce the
characteristic time
\begin{equation}
\label{eq4} \tau _\alpha = \left( {\frac{d^2}{2C_\alpha }}
\right)^{1 / \alpha }
\end{equation}
\noindent where d is the least distance to which the molecules can
approach to each other (if the molecules are considered as
spheres of the radius a then $d=2a$ \cite{Ab61}). For the
homogeneous case the normal diffusion coefficient $C_1$ is given
by the Stokes formula
\begin{equation}
\label{eq5} C_1 =\frac{k_B T}{6\pi a\eta}
\end{equation}
where $\eta$ the viscosity of the media and consequently the
characteristic time is
\begin{equation}
\label{eq6} \tau _1 = \frac{12\pi a^3\eta}{k_B T }
\end{equation}

The generalization of the ordinary diffusion equation within the
fractional calculus suggested by Schneider and Wyss
\cite{Sc89} is
\begin{equation}
\label{eq7} \frac{\partial P({\bf r},t)}{\partial t} = C_\alpha
(D_{0 + }^{1 - \alpha } \nabla^2 P)({\bf r},t)
\end{equation}
\noindent where $P({\bf r},t)$ is the probability density
function to find the particle at position ${\bf r}$ at time t,
$\nabla^2$ is the three-dimensional Laplace operator, C$_{\alpha
}$ denotes the fractional diffusion constant with the dimension
[cm$^{2}$/s$^{\alpha }$] and  D$^{1-\alpha }_{0 + }$ is the
Riemann-Liouville fractional derivative of order $1-\alpha $ and
with lower limit $0+$ which is defined via the following
relationship \cite{Ol74}
\begin{equation}
\label{eq8} \left( {D_{0 + }^{1-\alpha} f} \right)(x) =
\frac{1}{\Gamma (\alpha )}\frac{d}{dx}\int\limits_0^x {(x -
y)^{\alpha - 1}f(y)dy}
\end{equation}
\noindent where $\Gamma(x)$ is a gamma function. The solution of
the equation (\ref{eq7}) for the case of sub-diffusion $0 < \alpha
\le $ 1 with the initial condition $P({\bf r},0) = \delta ({\bf r})$
 where $\delta (x)$ is a Dirac function is obtained in
\cite{Sc89} and expressed via the Fox's function \cite{Sr82},
\cite{Pr86}, \cite{Gl93}
\begin{equation}
\label{eq9} P({\bf r},t) = \frac{1}{(r^2\pi )^{3 / 2}}H_{12}^{20}
\left( {\frac{r^2}{4C_\alpha t^\alpha }\left| {\begin{array}{l}
 (1,\alpha ) \\
 (3 / 2,1),(1,1) \\
 \end{array}} \right.} \right)
\end{equation}
\noindent The latter is defined as $H_{pq}^{mn} \left( {z\left|
{\begin{array}{l}
 (a_1 ,A_1 ),...,(a_p ,A_p ) \\
 (b_1 ,B_1 ),...,(b_p ,B_p ) \\
 \end{array}} \right.}
 \right)=\frac{1}{2\pi i}\int_L dsz^{-s}\eta(s)$
where
$\eta(s)=\frac{\prod_{i=1}^m\Gamma(b_i+B_is)\prod_{i=1}^n\Gamma(1-a_i-A_is)}
{\prod_{i=n+1}^p\Gamma(a_i+A_is)\prod_{i=m+1}^q\Gamma(1-b_i-B_is)}$.
The nomenclature in the Fox's function associated with the
vertical bar is explained via its explicit definition by the
contour integral. The requirements to the contour path $L$ are
formulated in \cite{Sr82}.

\section{Direct problem}

In this preliminary Sec. we solve the direct problem - how to
calculate the spectral density $J(\omega)$ knowing the FDC
 C$_{\alpha}$ ? - and present the detailed derivation of
 the result obtained in \cite{Sit05}.
We follow the algorithm of \cite{Ab61} (in what follows all
corresponding results from \cite{Ab61} are obtained as a
particular case $\alpha $=1 of the present approach). Under
 ${\bf r}$ we denote the vector ${\bf r}_1 - {\bf r}_2 $ connecting
two identical molecules diffusing relative to each other rather
than the radius-vector of the molecule diffusing relative to a
fixed point. This leads only to the change of $4C_\alpha t^\alpha
$ by $8C_\alpha t^\alpha $ in (\ref{eq6}). Our aim is to calculate
the correlation function
\begin{equation}
\label{eq10} G(t) = N\int\!\!\!\int {\frac{\Upsilon _2^{m^ \ast}
(\theta(0),\varphi(0))}{r_0^3 }} \frac{\Upsilon _2^m
(\theta(t),\varphi(t))}
{r^3}P({\bf r} - {\bf r}_0 ,t)d^3{\bf r}_0 d^3{\bf r}
\end{equation}
 where $N$ is the number of spins in 1 cm$^{3}$,
$\Upsilon _n^m(\theta,\varphi)$ is a spherical harmonic,
$P({\bf r},t)$ is given by (\ref{eq9}) and $\ast$
denotes complex conjugate. To be more precise we need the
spectral density of the correlation function $G(t)$ to calculate
the spin-lattice relaxation rate constant with the help of
(\ref{eq2}). At integration in (\ref{eq10}) one should take into
account that $r$ and $r_{0}$ can not be less than some limit
value $d$ -- the least distance to which the molecules can
approach to each other. If the molecules are considered as
spheres of the radius $a$ then $d=2a$ \cite{Ab61}.

First we make the Fourier transforming of the function
 $P({\bf r} - {\bf r}_0 ,t)$ with the mentioned above
 change of $4C_\alpha t^\alpha$ by $8C_\alpha t^\alpha $

\[
\frac{1}{\mid {\bf r} - {\bf r}_0 \mid^3}H_{12}^{20}
\left( {\frac{({\bf r} - {\bf r}_0)^2}
{8C_\alpha t^\alpha }\left| {\begin{array}{l}
 (1,\alpha ) \\
 (3 / 2,1),(1,1) \\
 \end{array}} \right.} \right)=
\]
\begin{equation}
\label{eq11}\frac{1}{(2\pi)^3}\int d^3 {\bf v} f({\bf v})
\exp\Bigl[i{\bf v}({\bf r} - {\bf r}_0)\Bigr]
\end{equation}
 Denoting
\begin{equation}
\label{eq12} {\bf R} = {\bf r} - {\bf r}_0;\ \ \ \ \ \ \ \ \ \
\ \ \ \ \ \ \ \ \
 R = \mid {\bf r} - {\bf r}_0 \mid;\ \ \ \ \ \ \ \ \ \ \ \ \ \ \
v  =\mid {\bf v} \mid
\end{equation}
we have the inverse transform (making use of spherical coordinates)
\begin{equation}
\label{eq13} f(v)=\frac{4 \pi}{v}\int\limits_0^\infty dR\
\frac{\sin \bigl( v R \bigr)}{R^2}H_{12}^{20}
\left( {\frac{R^2}
{8C_\alpha t^\alpha }\left| {\begin{array}{l}
 (1,\alpha ) \\
 (3 / 2,1),(1,1) \\
 \end{array}} \right.} \right)
\end{equation}
Substituting (\ref{eq13}) into (\ref{eq11}), then substituting
 the result into (\ref{eq10}) and making use of the known identities for
spherical functions
\begin{equation}
\label{eq14} \int d\Omega\
\Upsilon _l^{p^ \ast}(\Omega) \Upsilon _{l'}^{p'}(\Omega)=
\delta_{ll'}\delta_{pp'}
\end{equation}
\begin{equation}
\label{eq15} \exp(-i{\bf v}{\bf r})=4\pi
\Biggl(\frac {\pi}{2 v r}\Biggr)^{1/2} \sum_{p,l}i^l
 \Upsilon _l^{p^ \ast}(\Omega) \Upsilon _l^p(\Omega')
J_{l+1/2}(v r)
\end{equation}
 where $J_\nu(x)$ is a Bessel function of order $\nu$ we obtain
\begin{equation}
\label{eq16} G(t) =\frac{N}{\pi ^{3 / 2}}\int\limits_0^\infty
dv\ f(v)\Biggl[\int\limits_d^\infty dr\ \frac{J_{5/2}(vr)}{r^{3/2}}
\Biggr]^2
\end{equation}
where $f(v)$ is given by (\ref{eq13}). Taking the known value of
the inner integral
\begin{equation}
\label{eq17} \int\limits_d^\infty dr\ \frac{J_{5/2}(vr)}{r^{3/2}}=
v^{1/2}(vd)^{-3/2}J_{3/2}(vd)
\end{equation}
and denoting $u=vd$ we obtain
\[
G(t) = \frac{4N}{\pi ^{1 / 2}d^2}\int\limits_0^\infty
{\frac{du}{u^2}} \left[ {J_{\frac{3}{2}} (u)}
\right]^2\int\limits_0^\infty {\frac{dR}{R^2}} \sin \left(
{\frac{uR}{d}} \right)\times
\]
\begin{equation}
\label{eq18} H_{12}^{20} \left( {\frac{R^2}{8C_\alpha t^\alpha
}\left| {\begin{array}{l}
 (1,\alpha ) \\
 (3 / 2,1),(1,1) \\
 \end{array}} \right.} \right)
\end{equation}
The integration can be fulfilled and yields the expressions for both the
correlation function and its spectral density via Fox's functions.
However a Fox's function is to regret not tabulated at present
either in {\it Mathematica} or {\it Maple} or {\it Matlab}. Thus
it is rather difficult to use such formulas for plotting the
behavior of the correlation function and the spectral density.
That is why it is useful to obtain another representation of the
spectral density which enables one to plot the frequency
dependence of this function. For this purpose we make the Fourier
transforming of the Fox function in (\ref{eq18}). The latter can
be achieved with the help of the following trick going back to
original investigations of Fox. First we make use of (35) from
 the paper \cite{Sc89} to write
\begin{equation}
\label{eq19}  H_{12}^{20} \left( {\frac{R^2}{8C_\alpha t^\alpha
}\left| {\begin{array}{l}
 (1,\alpha ) \\
 (3 / 2,1),(1,1) \\
 \end{array}} \right.} \right)=\frac{1}{2}
 H_{12}^{20} \left( {\frac{R t^{-\alpha/2}}{(8C_\alpha)^{1/2}
}\left| {\begin{array}{l}
 (1,\alpha/2 ) \\
 (3 / 2,1/2),(1,1/2) \\
 \end{array}} \right.} \right)
\end{equation}
Then with the help of transform (57) $\leftrightarrow$ (56)
 from \cite{Wy99} we obtain
\[
M\left\{ \frac{1}{2}
{H_{12}^{20} \left( {\frac{R}
{(8C_\alpha \tau_{\alpha}^\alpha)^{1/2}}
\Biggl(\frac{t}{\tau_{\alpha}}\Biggr)^{-\alpha/2}\left|
{\begin{array}{l}
 (1,\alpha/2 ) \\
 (3 / 2,1/2),(1,1/2) \\
 \end{array}} \right.} \right)};s \right\} =
\]
\begin{equation}
\label{eq20} \frac{2^{-2s/\alpha}}{\alpha }
\left( {\frac{2R}{(8C_\alpha \tau_{\alpha}^\alpha )^{1 /
2}}} \right)^{\frac{2s}{\alpha }}\Gamma \left( {\frac{3}{2} -
\frac{s}{\alpha }} \right)\frac{\Gamma (1 - s / \alpha )}{\Gamma
(1 - s)}
\end{equation}
where $M\{...;s\}$ denotes the Mellin transform and
$\tau_\alpha$ is given by (\ref{eq4}).
We introduce the designations
\begin{equation}
\label{eq21} x=\frac{t}{\tau_\alpha};
\ \ \ \ \ \ \ \ \ \ \ \ \ \ z=\omega \tau_\alpha;
\ \ \ \ \ \ \ \ \ \ \ \ \ \
r=\frac{2R}{(8C_\alpha \tau_{\alpha}^\alpha )^{1 /
2}}
\end{equation}
and make use of the identity \cite{Sc95}
\begin{equation}
\label{eq22} F_C \left\{ {f(x);z } \right\} = M^{ - 1}\left\{
{\Gamma (s)\cos \frac{\pi s}{2}M\left\{ {f(x);1 - s} \right\}}
\right\}
\end{equation}
 where $F_{C}$ denotes the cosine Fourier transform and
$M^{-1}\{...\}$ denotes the inverse Mellin transform.
Thus for our function of
 interest
\begin{equation}
\label{eq23} f(x)=H_{12}^{20} \left( {\frac{R^2}
{8C_\alpha \tau^\alpha}x^{-\alpha}\left| {\begin{array}{l}
 (1,\alpha ) \\
 (3 / 2,1),(1,1) \\
 \end{array}} \right.} \right)
\end{equation}
we have
\begin{equation}
\label{eq24} M\{f(x);1-s\}=\frac{1}{\Gamma
(s)\alpha}
\Biggl(\frac{r}{2}\Biggr)^{\frac{2(1-s)}{\alpha}}
\Gamma \left( {\frac{3}{2} - \frac{1}{\alpha}+
\frac{s}{\alpha }} \right)
\Gamma \left(1 -\frac{1}{\alpha}+
\frac{ s }{ \alpha}\right)
\end{equation}
and
\begin{equation}
\label{eq25} F \left\{ {f(x);z } \right\} =
 2F_C \left\{ {f(x);z } \right\} =
\frac{2}{\alpha}\left(\frac{r}{2}\right)^{2/\alpha}
G(z)
\end{equation}
where
\begin{equation}
\label{eq26} G(z)=
M^{-1}\left\{\cos \frac{\pi s}{2}\left(\frac{r}{2}\right)^{-2s/\alpha}
\Gamma \left( {\frac{3}{2} - \frac{1}{\alpha}+
\frac{s}{\alpha }} \right)\Gamma \left(1 -\frac{1}{\alpha}+
\frac{ s }{ \alpha} \right)\right\}
\end{equation}
The calculation of the inverse Mellin transform in this formula
requires routine manipulations with the help of its
known properties \cite{Bat54}\\
\noindent $a^{-s}g(s)\leftrightarrow f(ax);\ \ \ \ \ \ \ \
g(s/h)\leftrightarrow hf(x^h);\ \ \ \ \ \ \ \
\Gamma (s)\leftrightarrow e^{-x}\\
g_1(s+A)g_2(s+A+B+1)\leftrightarrow x^A\int\limits_0^\infty
d\xi\ \xi^B f_1(x/\xi)f_2(\xi);\\
\sin(\alpha \pi s/2)\Gamma(s)
\leftrightarrow e^{-x\cos(\pi \alpha/2)}
\sin\left(x\sin(\alpha \pi /2)\right);\\
\cos(\alpha \pi s/2)\Gamma(s)
\leftrightarrow e^{-x\cos(\pi \alpha/2)}
\cos\left(x\sin(\alpha \pi /2)\right)$.
As a result of tedious but straightforward calculations we obtain
\[
G(z)=-\alpha
 \left(\left(\frac{r}{2}\right)^{2/\alpha}z\right)^{\alpha-1}\times
\]
\begin{equation}
\label{eq27}
\int_0^\infty d\xi \xi^{-1/2} \exp(-\xi)
\exp\left(-\frac{r^2z^\alpha}{4\xi}\cos \frac{\pi \alpha}{2}\right)
\sin\left\{\frac{r^2z^\alpha}{4\xi}\sin \frac{\pi \alpha}{2}-
\frac{\pi \alpha}{2}\right\}
\end{equation}
The substitution of the results into spectral density of the
correlation function (\ref{eq10}) yields
\[
J(\omega ) = -\frac{N\tau _\alpha^{1-\alpha}}
{\pi^{1/2} d^2 C_\alpha
 \left( {\omega \tau _\alpha } \right)^{1 - \alpha }}
\int\limits_0^\infty dR
\int\limits_0^\infty
\frac{du}{u^2} \left[ J_{\frac{3}{2}} (u)
\right]^2 \sin \left(\frac{uR}{d}\right)\times
\]
\[
\int_0^\infty d\xi \xi^{-1/2} \exp(-\xi)
\exp\left(-\frac{R^2(\omega \tau _\alpha)^\alpha}
{\xi 8C_\alpha \tau_{\alpha}^\alpha}\cos \frac{\pi \alpha}{2}\right)\times
\]
\begin{equation}
\label{eq28}\sin\left\{\frac{R^2(\omega \tau _\alpha)^\alpha}
{\xi 8C_\alpha \tau_{\alpha}^\alpha}\sin \frac{\pi \alpha}{2}-
\frac{\pi \alpha}{2}\right\}
\end{equation}
We denote
\begin{equation}
\label{eq29} a=\frac{R^2(\omega \tau _\alpha)^\alpha}
{8C_\alpha \tau_{\alpha}^\alpha}\sin \frac{\pi \alpha}{2}; \ \
b=\frac{\pi \alpha}{2};\ \ c=\frac{R^2(\omega \tau _\alpha)^\alpha}
{8C_\alpha \tau_{\alpha}^\alpha}\cos \frac{\pi \alpha}{2};\ \ \
x=\frac{1}{\xi}
\end{equation}
Then the inner integral over $\xi$ is
\[
I=\cos b \int_0^\infty dx\ x^{-3/2}
\exp\left(-\frac{1}{x}\right)\exp(-cx)\sin(ax)-
\]
\begin{equation}
\label{eq30}
\sin b \int_0^\infty dx\ x^{-3/2}
\exp\left(-\frac{1}{x}\right)\exp(-cx)\cos(ax)
\end{equation}
Both integrals here can be calculated with the help of N2.5.37.2 from
\cite{Pr81}. As a result of straightforward manipulations we obtain
\begin{equation}
\label{eq31} I=2g\left(\frac{\pi}{2h}\right)^{1/2}
\exp\left(-h\cos \frac{\pi \alpha}{4}\right)
\sin\left(h\sin \frac{\pi \alpha}{4}-\frac{\pi \alpha}{2}\right)
\end{equation}
where
\begin{equation}
\label{eq32}
g=\left[\frac{R^2(\omega \tau _\alpha)^\alpha}
{8C_\alpha \tau_{\alpha}^\alpha}\right]^{1/4};\ \ \ \ \ \ \
h=\frac{2R(\omega \tau _\alpha)^{\alpha/2}}
{(8C_\alpha \tau_{\alpha}^\alpha)^{1/2}}
\end{equation}
We denote
\begin{equation}
\label{eq33} B=\frac{2(\omega \tau _\alpha)^{\alpha/2}}
{(8C_\alpha \tau_{\alpha}^\alpha)^{1/2}}
\end{equation}
The integral over $R$ in (\ref{eq28})
\begin{equation}
\label{eq34} K=\int_0^\infty dR\ \sin \left(\frac{uR}{d}\right)
\exp\left(-BR\cos \frac{\pi \alpha}{4}\right)
\sin\left(BR\sin \frac{\pi \alpha}{4}-\frac{\pi \alpha}{2}\right)
\end{equation}
after simple manipulations can be cast into the combination of the
 integrals of the type N2.5.30.8
from \cite{Pr81}. As a result we obtain after straightforward
calculations
\begin{equation}
\label{eq35} K=-\frac{d u^3 \sin \frac{\pi \alpha}{2}}
{u^4 + d^2 u^2\frac{\left(\omega \tau
_\alpha\right)^{\alpha}}{C_\alpha \tau_{\alpha}^\alpha}
\cos \frac{\pi \alpha }{2}+\frac{d^4\left(\omega \tau
_\alpha\right)^{2\alpha}}{4C_\alpha^2 \tau_{\alpha}^{2\alpha}}}
\end{equation}
Taking into account the  definition (\ref{eq4}) we finally obtain
the formula for spectral density of the correlation function
in the general case of inhomogeneus media
($0<\alpha\le1$) \cite{Sit05}
\[
J(\omega ) = \frac{N\tau _\alpha ^{1 - \alpha } }{C_\alpha
d\left( {\omega \tau _\alpha } \right)^{1 - \alpha }}\sin
\frac{\pi \alpha }{2}\int\limits_0^\infty {du} \left[
{J_{\frac{3}{2}} (u)} \right]^2\times
\]
\begin{equation}
\label{eq36} \frac{u}{u^4 + \left(\omega \tau
_\alpha\right)^{2\alpha}+2u^2\left(\omega \tau
_\alpha\right)^{\alpha}\cos \frac{\pi \alpha }{2}}
\end{equation}
This formula generalizes that VIII.113 from \cite{Ab61} referring to a
particular case $\alpha=1$ to the case of arbitrary $0 < \alpha\le1$.
The integration in it can be easily done with the help of
{\it Mathematica} or {\it Maple} or {\it Matlab}. Thus this formula
is convenient for plotting the spectral density (see  \cite{Sit05}).
The substitution of this formula
into (\ref{eq1}),(\ref{eq2}),(\ref{eq3}) solves the direct
problem, i.e., how one can calculate the contribution into
spin-lattice relaxation time by anomalous translational diffusion
$(1/T_{1})_{trans}$ knowing the FDC $C_{\alpha }$?

\section{Inverse problem}
In this central Sec. we tackle some more difficult inverse problem,
i.e.,  how one can calculate the FDC $C_{\alpha }$ knowing the
 contribution
into spin-lattice relaxation time by anomalous translational
diffusion $(1/T_{1})_{trans}$? Our solution of this problem is
based on the fact that the integral in the (\ref{eq28}) can be cast
into a form of convolution for Mellin transform
 $\int_0^\infty \frac{du}{u}\varphi(u)K \left(\frac{x}{u}\right)
\leftrightarrow \tilde \varphi(s)\tilde K(s)$. We factorize the
characteristic time into the dimensional part $\tau_1$
($[\tau_1]=s$) and a dimensionless function $f(\alpha)$
($f(1)\equiv 1$ ) accounting for the dependence on the parameter
$\alpha$
\begin{equation}
\label{eq37} \tau _\alpha = \tau_1 f(\alpha)
\end{equation}
The function $f(\alpha)$ plays a key role in our analysis because
knowing it we have $\tau _\alpha$ and from (\ref{eq4}) we obtain
the desired value of the FDC $C_{\alpha }$. We denote the
dimensionless variable
\begin{equation}
\label{eq38} x = \omega_L \tau _1
\end{equation}
and introduce the function
\begin{equation}
\label{eq39} H(x,\alpha) = \frac{5d^3 \omega_L}{8\pi N\gamma^4
\hbar^2I(I+1)x}\left(\frac{1}{T_1}\right)_{trans}
\left(x,\alpha\right)
\end{equation}
We consider this function as
being extracted in principle from the experiment, i.e., as a
 primarily given
one. The condition of normalizing for the function $H(x,1)$ by the
requirement $f(1)\equiv 1$ will be given below. Making use of the
substitution $u=v^{\alpha/2}$ we obtain from (\ref{eq1}) and
(\ref{eq36})
\[
2xH(x,\alpha) = \alpha \left(f(\alpha)\right)^{\alpha}\sin
\frac{\pi \alpha }{2}\int\limits_0^\infty \frac{dv}{v} \left[
{J_{\frac{3}{2}} \left(v^{\alpha/2}\right)} \right]^2\times
\]
\[
\Biggl\lbrace
\frac{(x/v)^{\alpha}}{1+\left(x/v\right)^{2\alpha}
\left(f(\alpha)\right)^{2\alpha}+
2\left(x/v\right)^{\alpha}
\left(f(\alpha)\right)^{\alpha}\cos\frac{\pi\alpha}{2}}+
\]
\begin{equation}
\label{eq40} \frac{2^{\alpha
+1}(x/v)^{\alpha}}{1+\left(x/v\right)^{2\alpha}
\left(2f(\alpha)\right)^{2\alpha}+
2\left(x/v\right)^{\alpha}\left(2f(\alpha)\right)^{\alpha}
\cos\frac{\pi\alpha}{2}}\Biggr\rbrace
\end{equation}
Now we apply Mellin transform in the variable x (which we denote
 $M\{f(x),s\}=g(s)$ or $\leftrightarrow$) to both sides of
this equation. Making use of the property $x^\beta f(a x^h)
 \leftrightarrow h^{-1}a^{-(s+\beta)/h}g[(s+\beta)/h]$  where
 $a>0;\ h>0$ and of N6.8.33 from \cite{Bat54} we obtain
\begin{equation}
\label{eq41}  \left[
{J_{\frac{3}{2}} \left(x^{\alpha/2}\right)} \right]^2
\leftrightarrow \frac{2^{2s/\alpha}}{\alpha}
\frac{\Gamma \left (1-2s/\alpha \right)
\Gamma \left (3/2+s/\alpha \right)}
{\Gamma^2 \left (1-s/\alpha \right)
\Gamma \left (5/2-s/\alpha \right)}
\end{equation}
where $-3\alpha/2 < Re\ s < \alpha/2$. Making use of N6.2.12 from
\cite{Bat54} we obtain
\[
\frac{x^{\alpha}}{1+x^{2\alpha}
\left(f(\alpha)\right)^{2\alpha}+
2x^{\alpha}
\left(f(\alpha)\right)^{\alpha}\cos\frac{\pi\alpha}{2}}
\leftrightarrow
\]
\begin{equation}
\label{eq42}
\frac{\left(f(\alpha)\right)^{-(s+\alpha)}}{\alpha}
\frac{\pi\  \sin(\pi s/2)}{\sin(\pi \alpha/2) \sin(\pi s/\alpha)}
\end{equation}
where $-\alpha < Re\  s < \alpha$. Thus both transforms (\ref{eq41})
and (\ref{eq42}) have the common region
\begin{equation}
\label{eq43} -\alpha < Re\  s <\alpha/2
\end{equation}
For (\ref{eq42}) to be valid we must be sure
that
\begin{equation}
\label{eq44} \left(f(\alpha)\right)^{\alpha} > 0
\end{equation}
 To verify the latter
we obtain from (\ref{eq4}) and (\ref{eq37})
\begin{equation}
\label{eq45}\left(f(\alpha)\right)^{\alpha}=
\left( \frac{2C_1}{d^2}\right)^{ \alpha }
\left( \frac{d^2}{2C_\alpha }\right)
\end{equation}
We see that at $C_{\alpha}$  be a decreasing function with the decrese of
$\alpha$ from the value $\alpha =1$ (that is an inherent feature of
the subdiffusion) the function
$\left(f(\alpha)\right)^{\alpha}$ is an increasing one from the value
$\left(f(\alpha=1)\right)^{\alpha=1}=1$, i.e., indeed (\ref{eq44}) takes
place.

Applying (\ref{eq41}) and (\ref{eq42}) to the convolutions in (\ref{eq40})
we obtain
\[
2M\left\{xH(x,\alpha),s \right\}=
\frac{2^{2s/\alpha}}{\alpha}
\frac{\Gamma \left (1-2s/\alpha \right)
\Gamma \left (3/2+s/\alpha \right)}
{\Gamma^2 \left (1-s/\alpha \right)
\Gamma \left (5/2-s/\alpha \right)}\times
\]
\begin{equation}
\label{eq46} \frac{\pi\ \left(f(\alpha)\right)^{-s}
 \sin(\pi s/2)\left(1+2^{1-s}\right)}
{\sin(\pi s/\alpha)}
\end{equation}
Making use of the properties of a $\Gamma$-function we can obtain after
straightforward calculations
\begin{equation}
\label{eq47} \frac{\Gamma^2 \left (1-s/\alpha \right)
\Gamma \left (5/2-s/\alpha \right)}
{\Gamma \left (1-2s/\alpha \right)
\Gamma \left (3/2+s/\alpha \right)}=
\frac{\pi 2^{4s/\alpha-1}\left(3/2-s/\alpha\right)
\left(1/2-s/\alpha\right)}
{\left(1/2+s/\alpha\right)\sin\left( \frac{\pi s
}{2}\right)\Gamma\left(2s/\alpha\right)}
\end{equation}
Substituting it into (\ref{eq45}) we obtain
\[
\sin\left( \frac{\pi s
}{2}\right)\Gamma\left(2s/\alpha\right)
\left(2^{2/\alpha}f(\alpha)\right)^{-s}=
\]
\begin{equation}
\label{eq48}\alpha
M\left\{xH(x,\alpha);s\right\}
\frac{\left(3/2-s/\alpha\right)\left(1/2-s/\alpha\right)}
{\left(1/2+s/\alpha\right)\left(1+2^{1-s}\right)}
\end{equation}
With the help of N7.3.9 from \cite{Bat54} we obtain
\[
 \sin\left( \frac{\pi s
}{2}\right)\Gamma\left(2s/\alpha\right)
\left(2^{2/\alpha}f(\alpha)\right)^{-s}\leftrightarrow
\]
\begin{equation}
\label{eq49}
\frac{\alpha}{2}\exp\left[-2\left(xf(\alpha)\right)^{\alpha/2}
\cos\left( \frac{\pi \alpha
}{4}\right)\right]
\sin\left( 2\left(xf(\alpha)\right)^{\alpha/2}
\sin\left(\frac{\pi \alpha}{4}\right)\right)
\end{equation}
where $Re\ s > -\alpha/2 $. Thus the common region for all
transforms is
\begin{equation}
\label{eq50} -\alpha/2 < Re\  s <\alpha/2
\end{equation}
Taking the inverse Mellin transform for both sides of (\ref{eq48})
 we obtain
\[
\frac{1}{2}exp\left[-2\left(xf(\alpha)\right)^{\alpha/2}
\cos\frac{\pi\alpha}{4}\right]
\sin\left[2\left(xf(\alpha)\right)^{\alpha/2}
\sin\frac{\pi\alpha}{4}\right]=
\]
\begin{equation}
\label{eq51}\int\limits_0^\infty
du H(u,\alpha)G\left(\frac{x}{u}\right)
\end{equation}
where the fuction $G(x)$ is defined as the inverse Mellin transform
\begin{equation}
\label{eq52} G(x) \leftrightarrow \frac{\left(3/2-s/\alpha\right)
\left(1/2-s/\alpha\right)}
{\left(1/2+s/\alpha\right)\left(1+2^{1-s}\right)}
\end{equation}
The poles of its righthand side are $s=-\alpha/2$ and the solutions of the
equation $1+2^{1-s}=0$\  that are given by
 $s_m=1-\frac{i\pi}{ln 2}(2m+1)$ where $m=0;\pm1;\pm2;\ ...$.

Further we consider the case
\begin{equation}
\label{eq53} x < 1
\end{equation}
 that is sufficient for all practical experimental situations.
Then at calculation of the integral
\begin{equation}
\label{eq54} G(x)=\frac{1}{2\pi i}\int_{\sigma-i\infty}^{\sigma+i\infty}
ds\ x^{-s}\frac{\left(3/2-s/\alpha\right)
\left(1/2-s/\alpha\right)}
{\left(1/2+s/\alpha\right)\left(1+2^{1-s}\right)}
\end{equation}
we can choose $0 < \sigma < 1/2$ and close the contour in the lefthand
halfplane (due to (\ref{eq53})). All the poles $s_m$ appear to be
beyond its interior and only the pole $s=-\alpha/2$ is inside it. Thus
\begin{equation}
\label{eq55} G(x)=res\left\{
x^{-s}\frac{\left(3/2-s/\alpha\right)
\left(1/2-s/\alpha\right)}
{\left(1/2+s/\alpha\right)
\left(1+2^{1-s}\right)}\right\}_{s=-\alpha/2}=
\frac{2x^{\alpha/2}}{1+2^{1+\alpha/2}}
\end{equation}
 As a result we obtain
\[
exp\left[-2\left(xf(\alpha)\right)^{\alpha/2}
\cos\frac{\pi\alpha}{4}\right]
\sin\left[2\left(xf(\alpha)\right)^{\alpha/2}
\sin\frac{\pi\alpha}{4}\right]=
\]
\begin{equation}
\label{eq56}\frac{4x^{\alpha/2}}{1+2^{1+\alpha/2}}\int\limits_0^\infty
du H(u,\alpha)u^{-\alpha/2}
\end{equation}
This equation implicitly defines the required function
$f(\alpha)$ via the given function $H(x,\alpha)$. It is useful to
cast it into another form making use of the fact that the left
side of this equation has the form of the producing function for a
Chebyshev polynom \cite{Ang50}
\[
-\sum^\infty_{n=1}\frac{\left(-2\left(xf(\alpha)\right)^{\alpha/2}
\right)^n}{n!}
U_n\left(\cos\frac{\pi \alpha}{4}\right)=
\]
\begin{equation}
\label{eq57}\frac{4x^{\alpha/2}}{1+2^{1+\alpha/2}}\int\limits_0^\infty
du H(u,\alpha)u^{-\alpha/2}
\end{equation}

Now we take into account that in all practically important
situations we can adopt the requirement
\begin{equation}
\label{eq58} x << 1
\end{equation}
(e.g., in \cite{Sit05} we had $x = \omega_L \tau _1=0.015$).
In this case we can restrict
ourselves by the first term in the sum only. Taking into account
that $U_1\left(\cos\frac{\pi \alpha}{4}\right)=\sin\frac{\pi
\alpha}{4}$ \cite{Ang50} we finally obtain
\begin{equation}
\label{eq59} f(\alpha)\approx \left[\frac{2}{\sin\left(\frac{\pi
\alpha
}{4}\right)\left(1+2^{1+\alpha/2}\right)}\int\limits_0^\infty du
H(u,\alpha)u^{-\alpha/2}\right]^{2/\alpha}
\end{equation}
From the requirement $f(1)\equiv 1$ we obtain the condition of
normalizing for the function $H(x,1)$
\begin{equation}
\label{eq60} \int\limits_0^\infty du H(u,1)u^{-1/2}\approx
1+2^{-3/2}
\end{equation}
The desired dependence of the FDC on the parameter $\alpha$ hence has
the form
\begin{equation}
\label{eq61} C_{\alpha}\approx
C_1\tau_1^{1-\alpha}\left[\frac{2}{\sin\left(\frac{\pi \alpha
}{4}\right)\left(1+2^{1+\alpha/2}\right)}\int\limits_0^\infty du
H(u,\alpha)u^{-\alpha/2}\right]^{-2}
\end{equation}
This formula is the central result of this chapter.

\section{NMR relaxation or PFG NMR?}
As is well known there are  two strategies to employ NMR for measuring
translational diffusion coefficient: the NMR relaxation method
\cite{Ab61} and the PFG NMR (for review see \cite{Pri97}, \cite{Pri98}
 and refs. therein). Both of them have their merits and limitations nicely
discussed in \cite{Pri97}. As a matter of fact in the realm of ordinary
diffusion the PFG NMR is a fine Prince while the NMR relaxation is a
 Beggar. However their roles are not so obvious for the case of anomalous
diffusion.

The theory of PFG NMR is developed on the base of a combination of Bloch
equations with an ordinary diffusion term pioneered by Torrey
\cite{Tor56}, \cite{Sli80}. The scheme
works well for the case of unrestricted isotropic diffusion but becomes
analytically intractable for the case of anisotropic diffusion or that
within a confined geometry. While simple geometries are still amenable to
analytical approximations more complicated ones require numerical
solutions \cite{Pri97}. It is rather problematic to apply this scheme to
anomalous diffusion in complex media (in the sense of \cite{Wes03},
\cite{Met00}, \cite{Met04}) where even the
geometry as a rule can not be defined.

It is tempting to develop the theory of PFG NMR suitable for the description
of the anomalous diffusion along the line of the papers \cite{Kim02},
\cite{Kl02}, \cite{Sit05} and this chapter. However such a project also
encounters severe difficulties. To ehxibit their origin we briefly
scetch the mathematical scheme referring for explanation of designations to
,e.g., \cite{Sli80}. In our case instead of (G.7) from \cite{Sli80}
 we have a fractional Torrey equation
\[
\frac{\partial M^{+}({\bf r},t)}{\partial t} =
-i\gamma z\left(\frac{\partial H}{\partial z}\right)M^{+}({\bf r},t)-
\frac{M^{+}({\bf r},t)}{T_2}+
\]
\begin{equation}
\label{eq62}
C_\alpha
(D_{0 + }^{1 - \alpha } \nabla^2 M^{+})({\bf r},t)
\end{equation}
Analogously (G.9) from \cite{Sli80} we seek the solution in the form
\begin{equation}
\label{eq63}  M^{+}({\bf r},t) = M_0 \exp \left(-\frac{t}{T_2}\right)
\exp \left[-i\gamma z \left(\frac{\partial H}
{\partial z}\right)t\right]A(t)
\end{equation}
Then for the function $A(t)$ we obtain a fractional differential equation
\[
 \exp \left(-\frac{t}{T_2}\right)
\exp \left[-i\gamma z \left(\frac{\partial H}
{\partial z}\right)t\right]\frac{dA(t)}{dt}=
\]
\begin{equation}
\label{eq64}
-C_\alpha \left(\gamma\frac{\partial H}
{\partial z}\right)^2 D_{0 + }^{1 - \alpha }\left\{t^2
\exp \left(-\frac{t}{T_2}\right)
\exp \left[-i\gamma z \left(\frac{\partial H}
{\partial z}\right)t\right]A(t)\right\}
\end{equation}
Recalling the definition of the fractional derivative (\ref{eq8})
 we can rewrite (\ref{eq64}) as an integro-differential equation
\[
 \exp \left(-\frac{t}{T_2}\right)
\exp \left[-i\gamma z \left(\frac{\partial H}
{\partial z}\right)t\right]\frac{dA(t)}{dt}=
-C_\alpha \left(\gamma\frac{\partial H}
{\partial z}\right)^2\times
\]
\begin{equation}
\label{eq65}
\frac{1}{\Gamma (\alpha )}\frac{d}{dt}\int\limits_0^t ds{(t -
s)^{\alpha - 1}s^2
\exp \left(-\frac{s}{T_2}\right)
\exp \left[-i\gamma z \left(\frac{\partial H}
{\partial z}\right)s\right]A(s)}
\end{equation}
We denote
\begin{equation}
\label{eq66} B=C_\alpha\left(\gamma\frac{\partial H}
{\partial z}\right)^2
\end{equation}
\begin{equation}
\label{eq67} c=\frac{1}{T_2}+
i\gamma z\left(\frac{\partial H}{\partial z}\right)
\end{equation}
and make the Laplace transform of (\ref{eq65}). Denoting
the Laplace transform of the function $A(t)$ as $\beta(p)$
\begin{equation}
\label{eq68} A(t) \leftrightarrow \beta(p)
\end{equation}
we obtain after standard manipulations
\begin{equation}
\label{eq69} (p+c)\beta(p+c)-A(0)=-B p^{1-\alpha}
\frac{d^2 \beta(p+c)}{dp^2}
\end{equation}
Introducing a new complex variable
\begin{equation}
\label{eq70} z=p+c
\end{equation}
we obtain an ordinary differential equation
\begin{equation}
\label{eq71} B(z-c)^{1-\alpha}\frac{d^2\beta(z)}{dz^2}+
z\beta(z)=A(0)
\end{equation}
The problem we encouter with is to solve the homogeneous part of the
 equation (\ref{eq71}). Introducing a new function $u(z)$ with the
 help of the relationship
\begin{equation}
\label{eq72} \frac{d\beta(z)}{dz}=\beta(z)u(z)
\end{equation}
we can cast the homogeneous part of the  equation (\ref{eq71}) into an equivalent
 form
\begin{equation}
\label{eq73} \frac{du(z)}{dz}+u^2(z)=-\frac{z}{B(z-c)^{1-\alpha}}
\end{equation}
The latter is a Riccati equation. As is well known there is no general
way to obtain analytical solution of the Riccati equation and the
author of this chapter failed to obtain the solution for our
 particular case. In our opinion the approach encounters severe mathematical
difficulties. We conclude that for the case of anomalous
diffusion in complex media the NMR relaxation method has advantage
over the PFG NMR because its mathemitical scheme is analytically
tractable as the previous Sec. shows. The latter conclusion should be
reconsidered because since this Chapter was submitted the problem of
fractional Torrey equation has been solved in \cite{Mag08}. In the next Sec.
we discuss some difficulties of
physical character that still remain and prevent the NMR relaxation method to
a become flawlessly accomplished one.

\section{Conclusion}
As is well known the main problem of the NMR relaxation method is that
"the relaxation mechanism
of the probe species needs to be known, and it is
required that the intermolecular contributions to
the relaxation can be separated from the intramolecular
 contributions" (see \cite{Pri97} and refs. therein). As for the first part
of the above argument the dipole-dipole relaxation mechanism dominates in
very many cases and can be reliably supported upon. However the second part
is really a problem. To put it in other words there are contributions to the
spin-lattice relaxation rate constant from both translational and rotational
diffusion of the tracer molecule
\begin{equation}
\label{eq74} (1/T_{1})=(1/T_{1})_{trans}+(1/T_{1})_{rot}
\end{equation}
The input information for the approach developed in this chapter is
$(1/T_{1})_{trans}$ and the problem is how to separate it reliably from the
rotationl contribution, i.e., to extract it from the experimentally
 observable
value $(1/T_{1})$. For the case of ordinary diffusion (both rotational and
translational) the problem is solved somehow because we know an explicit
 expression for
$(1/T_{1})_{rot}$ \cite{Ab61}. However in a complex media both
 translational and rotational contributions generally become
some functions of the
parameter $\alpha$ characterising the extent of inhomogeneity. While the
dependence $(1/T_{1})_{trans}(\alpha)$ can be calculated explicitly (see
\cite{Sit05} and Sec.3 in this chapter) the
dependence $(1/T_{1})_{rot}(\alpha)$ remains unknown.
To develop mathematically the approach for calculation of the latter
 dependence
along the line of \cite{Ab61} basing upon the results of the paper
\cite{Ayd04} is quite feasible. However a physical problem still remains:
 whether the
parameter $\alpha$ figuring in the fractional derivative of the
 equation for the
translational diffusion should be the same as that for the rotational one?
If they differ from each other what is the relationship between them?
The author of this chapter does not know the answers to these questions
at present. Untill this problem is resolved one should resort to some
 intuitive physical arguments. In our opinion it seems reasonable to assume
 that the  rotational diffusion
depends on the extent of inhomogeneity in a much narrow range than the
translational one. For instance in a porous media the translational
diffusion is very sensitive to the radius of the pores in the whole range of
this parameter while the rotational diffusion seems to become sensitive to it only
when the radius of the pores becomes commensurable with the radius of the
tracer molecule. If it is really so then in a wide range of the extent of inhomogeneity
$\alpha_c < \alpha \leq 1$ one can assume that $(1/T_{1})_{rot}$ is
independent of $\alpha$ and use the expression for it from \cite{Ab61}
 obtained for ordinary rotational diffusion. This trick enanles one to
extract the required value $(1/T_{1})_{trans}(\alpha)$ from the
experimentally observable
 one $(1/T_{1})(\alpha)$
\begin{equation}
\label{eq75} (1/T_{1})_{trans}(\alpha)=(1/T_{1})(\alpha)-
(1/T_{1})_{rot}
\end{equation}
The result of such procedure is the input information for the approach developed in this
chapter. If it is obtained then the below described strategy is feasible.

The formula (\ref{eq61}) relates the FDC with the function $H(x,\alpha)$ (\ref{eq39}).
 The latter
takes into account the  contribution to the spin-lattice
relaxation rate constant by anomalous translational diffusion
and can in principle be extracted from the experiment as is
described above. Besides $\alpha$ it depends on the
dimentionless parameter $x = \omega_L \tau _1$. In practice it is rather
problematic to vary the Larmor frequency $\omega_L$. On the other hand the
characteristic time (\ref{eq6})
$\tau _1 = \frac{12\pi a^3\eta}{k_B T }$ depends on temperature and can be
easily varied during the experiment. This fact enables one to retrieve the
dependence of the contribution to the spin-lattice relaxation time due
 to translational diffusion  $\left({T_1}\right)_{trans}
\left(x,\alpha\right)$ on $x$ at different $\alpha$ from
the experimental measurements. Approximation of the data obtained by a
suitable analytical function of two variables yields $ H(x,\alpha)$
and substitution of the
latter into (\ref{eq61}) yields the required dependence of the FDC on the
parameter $\alpha$.

We conclude that
 the formula (\ref{eq61}) is an ingredient in solving
 the problem - how to retrieve the
fractional the FDC from NMR relaxation data? The mathematical scheme
for NMR relaxation method is
analytically tractable in contrast to that of PFG NMR. In the latter
 case the attempts to
develop a theory suitable for the description of the anomalous
diffusion encouter severe mathematical difficulties. The result
obtained in this chapter  promotes the NMR
 relaxation method to become a powerful tool in solving the problem
of experimental measuring the fractional diffusion coefficient.
However much work still remains to be done for reliable separating the
contribution of translational diffusion into the spin-lattice relaxation
 rate constant from that of rotational diffusion in the experimentally
measured values. This preliminary  step is necessary for obtaining
input information for the approach developed in this chapter. \\

Acknowledgements. The author is grateful to Prof. A.V. Anisimov for
helpful discussions. The work was supported by the grant from
RFBR.

\end{document}